\documentclass[12pt]{spieman} 

\usepackage{amsmath,amsfonts,amssymb}
\usepackage{wrapfig}
\usepackage{setspace}
\usepackage{tocloft}
\usepackage{graphicx}
\usepackage{color,soul}
\graphicspath{{./}{figures/}}
\usepackage[colorlinks=true, allcolors=blue]{hyperref}
\newcommand\asec{\mbox{$^{\prime\prime}$}}%
\newcommand\lya{\mbox{Ly$\alpha$}}%

\title{Inflight performance and future improvements for the INtegral Field Ultraviolet Spectrographic Experiment, the first far ultraviolet integral field spectrograph}

\author[a,*]{Alex Haughton}
\author[b]{Emily M. Witt}
\author[a]{Brian T. Fleming}
\author[a]{Alex Sico}
\author[a]{Michael J. Kaiser}
\author[c]{M.S. Oey}
\author[a]{Grace Halferty}
\author[a]{Dmitry Vorobiev}
\author[a]{Kevin France}
\author[d]{Takashi Sukegawa}
\author[e]{Oswald Siegmund}
\author[e]{Adrian Martin}

\affil[a]{University of Colorado, Boulder, Laboratory for Atmospheric and Space Physics, 1234 Innovation Drive, Boulder, CO, USA, 80303}
\affil[b]{Johns Hopkins University, Department of Physics and Astronomy, 3400 N. Charles St, Baltimore, MD, USA, 21218}
\affil[c]{University of Michigan, Department of Astronomy, 323 West Hall, 1085 S. University Ave, Ann Arbor, MI, USA, 48109}
\affil[d]{Canon, Inc., Optical Products Operations, 20-2 Kiyoharakogyodanchi, Utsunomiya-Shi, Tochigi, Japan, 321-3231}
\affil[e]{Sensor Sciences, LLC, 3333 Vincent Road, Suite 103, Pleasant Hill, CA, USA, 94523}

\begin{document}
\maketitle

\begin{abstract}
Integral field spectroscopy allows for spectral mapping of extended sources in a time efficient manner. An integral field unit (IFU) in the ultraviolet on Habitable Worlds Observatory (HWO) could be used to quickly map extended objects like supernova remnants or galaxies and their surroundings, but there are technical challenges to an ultraviolet IFU. The INtegral Field Ultraviolet Spectrographic Experiment (INFUSE), a sounding rocket project, is the first static configuration far ultraviolet integral field spectrograph. INFUSE features an \textit{f}/16, 0.49m Cassegrain telescope and a 26-element image slicer feeding 26 replica holographic gratings, with spectra imaged by the largest cross-strip microchannel plate detector flown in space. The first launch of INFUSE occurred from White Sands Missile Range on October 29th, 2023, and demonstrated spectral multiplexing, successfully detecting ionizing gas emission in the XA region of the Cygnus Loop. INFUSE will launch again in fall 2025 to observe NGC 2366, a local analog for Green Pea type galaxies, with several enhancements including a xenon-enhanced lithium fluoride + aluminum coated grating, testing the leading flight coating for HWO for the first time. The INFUSE IFU is designed as a pathfinder for a potential IFU mode on HWO, enabling rapid 3D spectroscopy of extended sources. 
\end{abstract}


{\noindent \footnotesize\textbf{*}Alex Haughton, \linkable{alex.haughton@lasp.colorado.edu}}


\section{Introduction}
\label{sec:intro}  
Habitable Worlds Observatory (HWO) will build on the legacy of other flagship missions, leveraging technological advancements in ultraviolet (UV) instrumentation to be a successor to Hubble Space Telescope (HST)\cite{hwosite}. Regardless of where trade space discussions end, it will have the largest primary mirror of any UV telescope to date with the ability to ``study the universe with unprecedented sensitivity and resolution."\cite{nasahwo} The science goals of HWO span the decadal survey from Worlds and Suns in Context to Cosmic Ecosystems; considering the oversubscription rate of HST\cite{watkins23} and the budgets of flagship projects\cite{greenfield21}, observation time on HWO will be precious. Theoretical designs of HWO include a multi-object spectrograph (MOS) with a $\sim$ 2' x 2' field of view (FOV) with grating modes of multiple resolutions spanning the 100-400 nm bandpass\cite{clampin24}. This multiplexing device, likely an advanced version of the James Webb Space Telescope - Near InfraRed Spectrograph (JWST-NIRSpec) microshutter array\cite{luvoir,ferruit22}, enables simultaneous spectroscopy of hundreds of objects at a time but is limited to a single object per dispersion direction row of the device to avoid spectral confusion. 

The INtegral Field Ultraviolet Spectroscopic Experiment (INFUSE) overcomes this limitation of a MOS, recording spatially resolved spectra of a two-dimensional FOV in a single observation. INFUSE is designed as a pathfinder for a potential integral field spectrograph (IFS) mode on HWO, which would be a compact, likely static (single or limited-mode) compliment to the larger, multi-mode MOS.

This paper covers the performance of and improvements to INFUSE, focusing on progress and technologies relevant to HWO. The introduction will provide context to the development and use of integral field units (IFUs) and discuss some of the science enabled by this new observing capability. \S 2 then covers systems of INFUSE relevant to HWO and improvements to INFUSE for the second flight. \S 3 introduces the Spectroscopic Ultraviolet Multi-Object Observatory (SUMO), a ride along experiment on the second flight that will test digital micromirror devices (DMDs) in space for the first time. For more details on the design of INFUSE, please see ``INFUSE: preflight performance of a rocket-borne FUV integral field spectrograph" by Witt et al\cite{witt23} and ``INFUSE: inflight performance and future improvements for the first FUV integral field spectrograph to study the influence of massive stars on galaxies" by Haughton et al\cite{haughton24}.

\subsection{Integral field spectroscopy}
\label{sec:ifs}

\begin{figure}
\begin{center}
\includegraphics[width=3.5in]{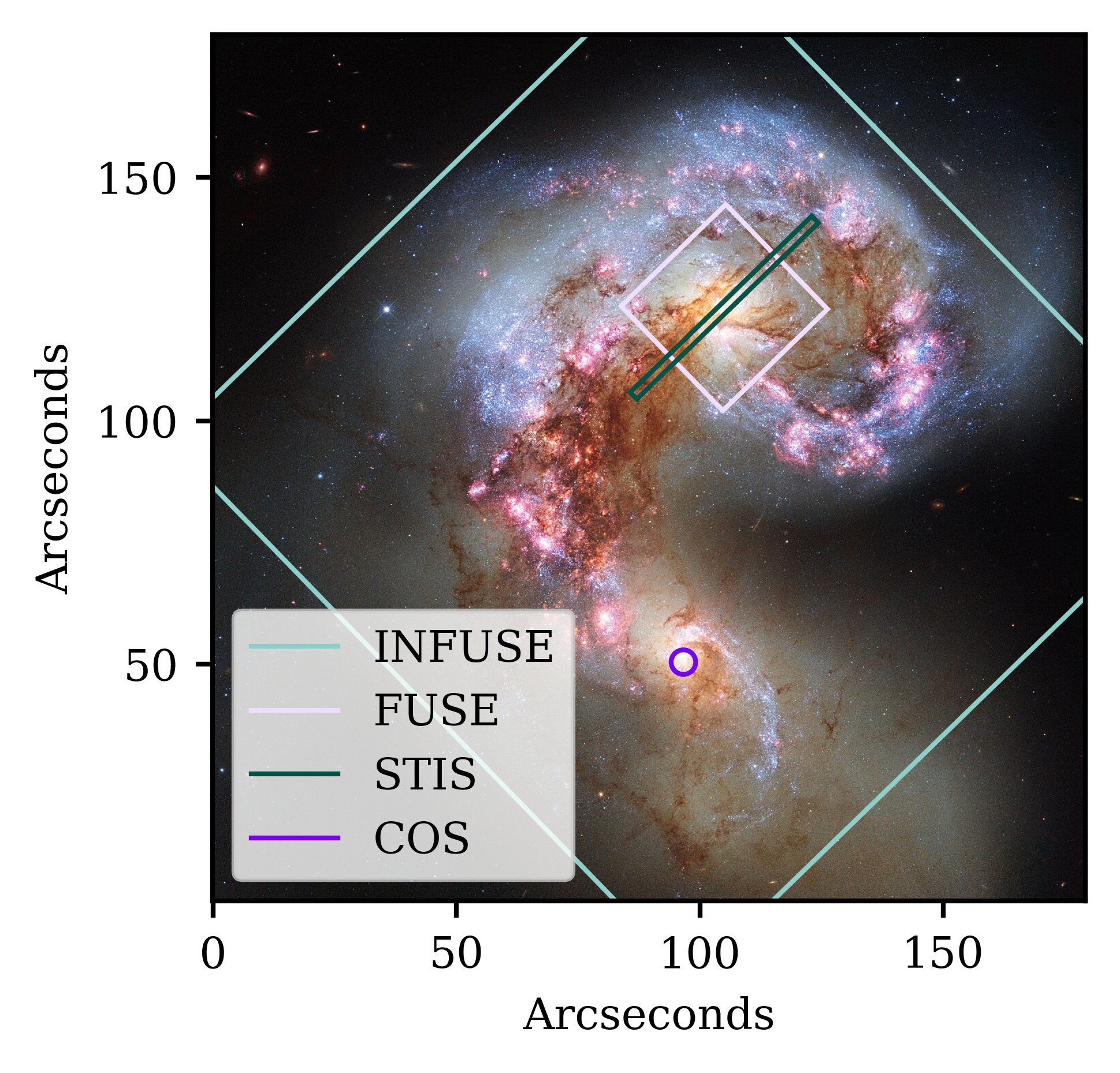}
\caption{The FOV of various UV spectrographs overlaid on an HST image of the Antennae galaxies \cite{whitemore06}. The Antennae are chosen as they are well matched to the INFUSE FOV for demonstration purposes and are well known galaxies that a potential future surveyor might observe; however, they are not currently INFUSE targets.}
\label{fig:fovintro}
\end{center}
\end{figure}

Integral field spectroscopy offers greater efficiency for comprehensive mapping of extended objects relative to long-slit or point source spectrographs. While point source spectrographs are sufficient for stars and other unresolved sources, and multi-object or long-slit spectrographs allow the observation of clusters of discrete objects or radial profiles of extended sources, these types of spectrographs require many pointings to fully map out broad sources such as supernova remnants or galaxies and their surroundings. IFSs of comparable sensitivity produce spectral maps of extended objects in significantly less time (Fig.~\ref{fig:fovintro}). As a competitively allocated resource like HST, time on HWO will be at a premium. An IFS mode on HWO will enable more efficient observation modes for galaxy studies, while also opening up the science opportunities outside the reasonable capability of a MOS.

In ground-based observing, IFUs such as the Multi-Unit Spectroscopic Explorer (MUSE) on the Very Large Telescope (VLT) \cite{bacon10} or the Faint Object Camera and Spectrograph on the Subaru Telescope \cite{ozaki20} have opened up new discovery spaces; MUSE is consistently the most productive VLT instrument\cite{muse}. In space, of the over 1900 hours of observing time awarded to study galaxies in JWST Cycle 2, the IFS mode on NIRSpec was used for $\sim$27\% of them\cite{cycle2}. Designing an UV IFU, however, presents additional technical challenges. Far ultraviolet (FUV) mirror reflectivity has traditionally been poor ($\sim$65-70\%)\cite{ohl00}, leading to loss of throughput, and UV observations must be made from space due to atmospheric losses, which adds size and weight restrictions, radiation and atomic oxygen concerns for electronics, launch costs, and the inability to readily service the instruments.

Prior to 2023, only the SPINR sounding rocket program and the Fireball balloon demonstrated UV IFUs\cite{cook03,milliard10}. However, SPINR used a traditional long slit scanner and mathematical deconvolution to disentangle spectral confusion, and Fireball was limited to wavelengths $>$ 190 nm and has since been re-scoped to a MOS to remove low throughput fiber optics. Viable designs of an IFU capable of covering the far and near UV are needed to inform the trade space for HWO\cite{clampin24}. INFUSE is the first true FUV IFU, capable of capturing over 1,000 resolved spectra spanning a 2.57' x 2.49' FOV in a single flight. INFUSE flew for the first time in October 2023, launching from White Sands Missile Range (WSMR) and observing the XA region of the Cygnus Loop supernova remnant. For its second flight in Fall 2025, INFUSE will study a local galaxy thought to be an analog for galaxies at the Epoch of Reionization (EoR).

\begin{figure}
\begin{center}
\includegraphics[width=\linewidth]{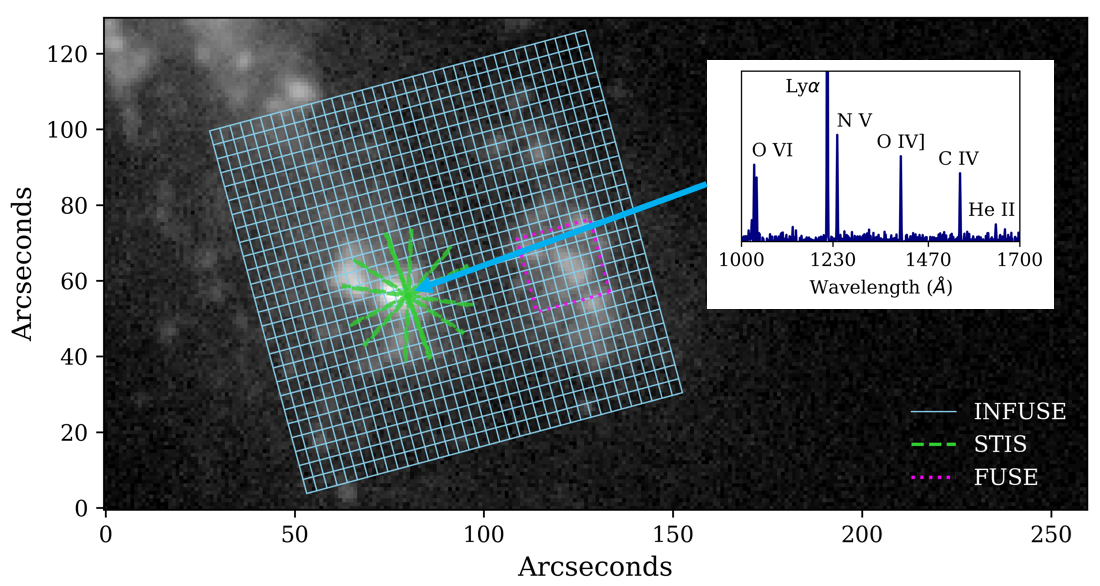}
\caption{FUV Galaxy Evolution Explorer image of NGC 2366 with companion galaxy NGC 2363. Each blue box of INFUSE represents an individual spectra that INFUSE will collect during its flight, with the dimensions set by the resolution of the instrument. The INFUSE pointing and angle was chosen to maximize the science return, focusing on Mrk 71 and the companion galaxy NGC 2363. The INFUSE flight provides greater coverage in the FUV than any previous observations.}
\label{fig:ngc2366}
\end{center}
\end{figure}

\subsection{UV integral field spectroscopy of NGC 2366}
\label{sec:scicase}


The intergalactic medium (IGM) of the universe was largely neutral before the EoR at redshift $\sim$ z $>$ 12, when photons thought to originate from young stars in early galaxies ionized the IGM. Direct measurement of ionizing photons from this era is not possible due to absorption by the IGM; therefore, studies have focused on lower redshift analogs \cite{bergvall06,leitet13,flury22}. A class of galaxies known as Green Peas (GPs) have proven to be particularly promising, frequently showing escape of ionizing photons as well as similarities with EoR galaxies observed by JWST \cite{rhoads23,mascia23}. Some of these traits include a high [OIII]/[OII] ratio, UV spectral slope $\beta < -2$, and high specific star formation rate \cite{flury22,amorin10}. Together, these are evidence of highly ionized starburst regions with hot young OB stars producing escaping ionizing photons. Unfortunately, GPs are too compact to resolve even with the largest ground-based observatories, meaning the mechanisms within the galaxies that drive these traits remain largely hidden.


\begin{figure}
\begin{center}
\includegraphics[width=\linewidth]{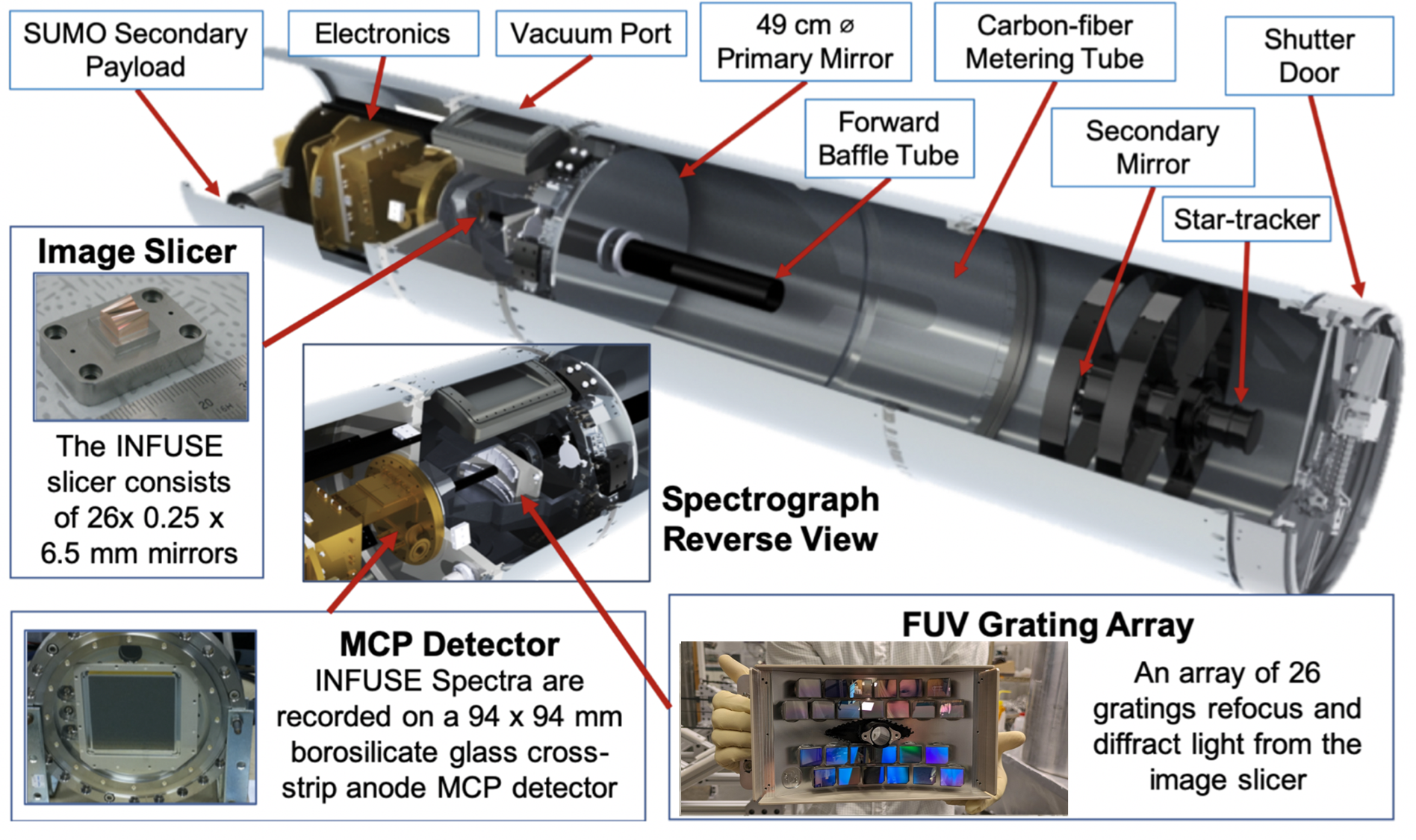}
\caption{Mechanical design of INFUSE, showing Cassegrain telescope, image slicer and grating array, MCP detector, and space for the SUMO secondary payload.     The payload as seen here is about 3 meters long.}
\label{fig:INFUSE}
\end{center}
\end{figure}

NGC 2366 is a nearby (3.4 Mpc) dwarf galaxy occupying about 5' x 3' on the sky that shares many of these GP traits, especially in the starburst region known as Markarian 71 (Mrk 71)\cite{micheva17}. Mrk 71 is a benchmark object for understanding massive star feedback at low metallicity where radiative cooling is expected to dominate\cite{oey23,oey17}. 3D data cubes (x, y, $\lambda$) generated from INFUSE observations of the FUV continuum and ionized and neutral line transitions will trace the star formation history and hot ionized gas distribution in the area. The O VI and C IV emission, tracers of hot ionized gas, will provide unparalleled diagnostic insight on the radiative cooling and energy budget of this landmark system. Knowledge gained from the INFUSE observation of NGC 2366 may increase understanding of the internal galactic mechanisms driving the escape of ionizing photons.

INFUSE will launch for a second flight on October 6, 2025 aboard a Black Brant IX sounding rocket to map this region in the FUV for the first time (Fig.~\ref{fig:ngc2366}). We require 300 seconds on target (capturing both Mrk 71 and the companion galaxy NGC 2363) to achieve an SNR greater than 5 depending on the region.

\section{The INtegral Field Ultraviolet Spectroscopic Experiment}
\label{sec:infuse}


The INFUSE instrument (Fig.~\ref{fig:INFUSE}) consists of an \textit{f}/16 Cassegrain telescope with a 490 mm parabolic primary mirror and a convex hyperbolic 110 mm diameter secondary mirror. The primary mirror mount has proven to be overconstrained and stresses the primary mirror, increasing the spot size of the instrument and limiting the spatial and spectral resolution. Light is focused onto an image slicer from Canon, Inc.\cite{suematsu17}. An image slicer is an optic consisting of a number of flat mirrors each tilted in a slightly different direction; each flat mirror acts as a long slit. The image slicer is a core enabling technology for INFUSE and was fabricated following the methodologies of Sukegawa et al\cite{sukegawa22}, with a micromachined invar substrate. The surface roughness of the two delivered image slicers are 7 and 11 \AA~RMS on the flight and spare units, respectively. 

The image slicer disperses light onto 26 replica holographic gratings from Horiba Jobin-Yvon, which diffract light onto a large format cross strip (XS) microchannel plate (MCP) detector from Sensor Sciences, LLC\cite{siegmund21}. The telescope, spectrograph, detector, and electronics are packaged in 22 inch sounding rocket skins, and launched on suborbital flights by the NASA Sounding Rockets Program. 

\begin{table}[ht]
\caption{Basic specifications of the INFUSE instrument. The designed instrument had higher spatial and spectral resolution, but an astigmatism was introduced by the primary mirror mount.} 
\label{tab:telspecs}
\begin{center}       
\begin{tabular}{|l|l|} 
\hline
\rule[-1ex]{0pt}{3.5ex}  \textbf{Specification} & \textbf{Measurement} \\
\hline
\rule[-1ex]{0pt}{3.5ex}  Primary size & 490 mm  \\
\hline
\rule[-1ex]{0pt}{3.5ex}  Field of view & 2.57' x 2.49' \\
\hline
\rule[-1ex]{0pt}{3.5ex}  Resolving power & $\sim$ 350 (filled aperture)\\
\hline
\rule[-1ex]{0pt}{3.5ex}  Spatial resolution & 4.18" - 4.83" x 5.7"\\
\hline
\rule[-1ex]{0pt}{3.5ex}  Bandpass & 1000 - 2000 \AA  \\
\hline 
\rule[-1ex]{0pt}{3.5ex}  Effective area & 130 cm$^{2}$ at 1130 \AA \\
\hline
\rule[-1ex]{0pt}{3.5ex}  Detector background & 0.18 cts/cm$^{2}$/s \\
\hline
\end{tabular}
\end{center}
\end{table}

INFUSE was launched from WSMR on October 29, 2023, and successfully demonstrated all of the technologies onboard in flight\cite{witt24}. While there were some issues caused by launch delays (\S \ref{sec:vac}), enough data was collected to observe C IV, O IV], and O VI in the Cygnus XA Loop (Fig.~\ref{fig:data}), surpassing spatial coverage from previous flights of the International Ultraviolet Explorer\cite{danforth01} and the Far Ultraviolet Spectroscopic Experiment (FUSE)\cite{wilkinson98}. A number of improvements are in progress or planned for the next flight: improved vacuum performance by replacing a feedthrough and polishing several o-ring seals, installation of a getter pump to help maintain a dry vacuum, increasing the FOV of the guide aspect camera that provides real time targeting, and roughening and redesign of spectrograph baffles to reduce scattered light. The second flight of INFUSE will also serve as the flight demonstration of a replica grating coated with aluminum (Al) and capped with xenon diflouride-enhanced lithium flouride (XeLiF)\cite{quijada22}.

\subsection{Detector development for HWO}
\label{sec:detect}
The detector is a 94 x 94 mm XS MCP detector, the largest ever flown with over 22 million resolution elements, and provides technical development for potential detectors on HWO\cite{clampin24}. The detector uses a solar-blind caesium iodide (CsI) photocathode, is photon-counting, and is digitized to 13-bit resolution for 12 micron digital pixels on INFUSE. Post-flight testing of the detector shows that standard UV-mission vacuum protocols to reduce water vapor exposure have kept the CsI from degrading, and the plates from the first flight will be used on the second flight. 

\begin{figure}
\begin{center}
\includegraphics[width=\linewidth]{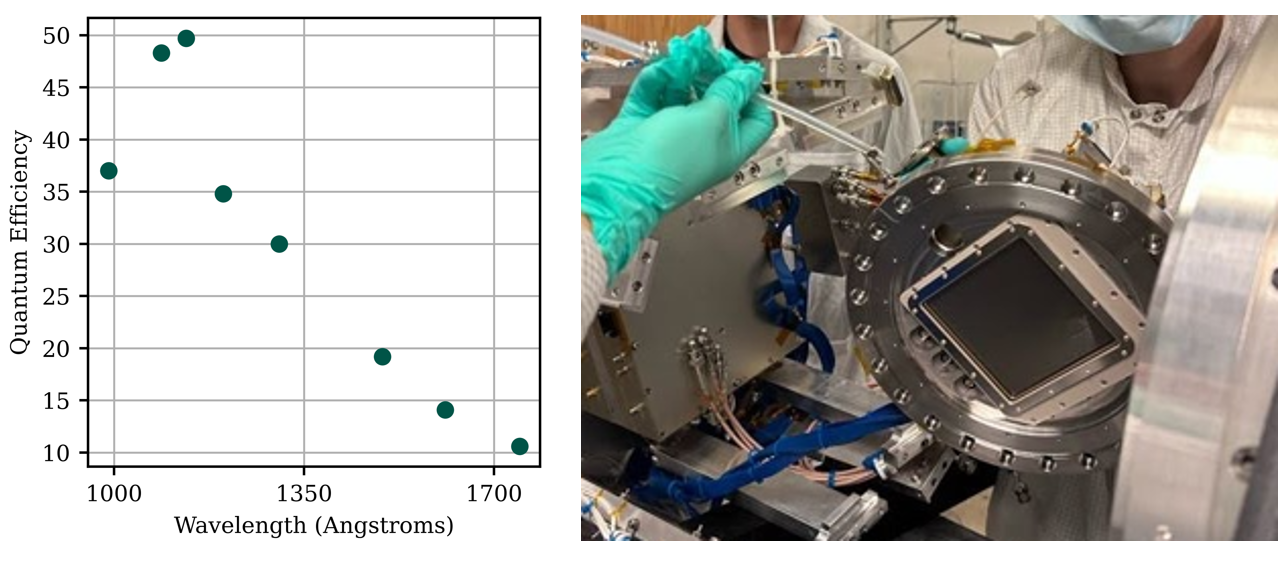}
\caption{Left: The quantum efficiency of the INFUSE detector, optimized for the O VI $\lambda\lambda$ 1032, 1036 doublet, peaks at near 50 percent. Right: INFUSE leverages the largest XS MCP detector ever flown.}
\label{fig:detect}
\end{center}
\end{figure}

The high quantum efficiency (49.7$\%$ at 1133 \r{A} (Fig.~\ref{fig:detect})) and low photon counting background (0.18 counts per cm$^{2}$ per second) of the INFUSE detector show MCPs are ideal for high sensitivity observations. In addition, due to the atomic layer deposition process, which provides a secondary emission layer of electrons, the INFUSE detector is resilient to the gain sag issues from cumulative events experienced by Cosmic Origins Spectrograph. The detector electronics output data via ethernet that is translated to multiplexed parallel data by custom interface electronics for telemetry to the ground.


\subsection{Vacuum improvements for second flight}
\label{sec:vac}

There were several events that occurred on launch night that led to a delay of over four hours, over which time INFUSE was not under active vacuum pumping. During this time the internal pressure rose to over 0.1 torr, exceeding the expected historical leak rate based on prior University of Colorado (CU) experiments.\cite{kane11,erickson21} As a result of this extended time at high pressure, gas was trapped in the INFUSE detector that took $\sim$350 seconds to be removed by the vacuum of space during flight. During this time, application of high voltage during flight ionized the trapped gas, resulting in a plasma that overloaded the current limiter on the detector. It was only after the outgassing into space reduced the pressure sufficiently that the detector activated, cutting the total observation time by approximately 300 seconds down to 19.4 seconds.

\begin{figure}
\begin{center}
\includegraphics[width=4in]{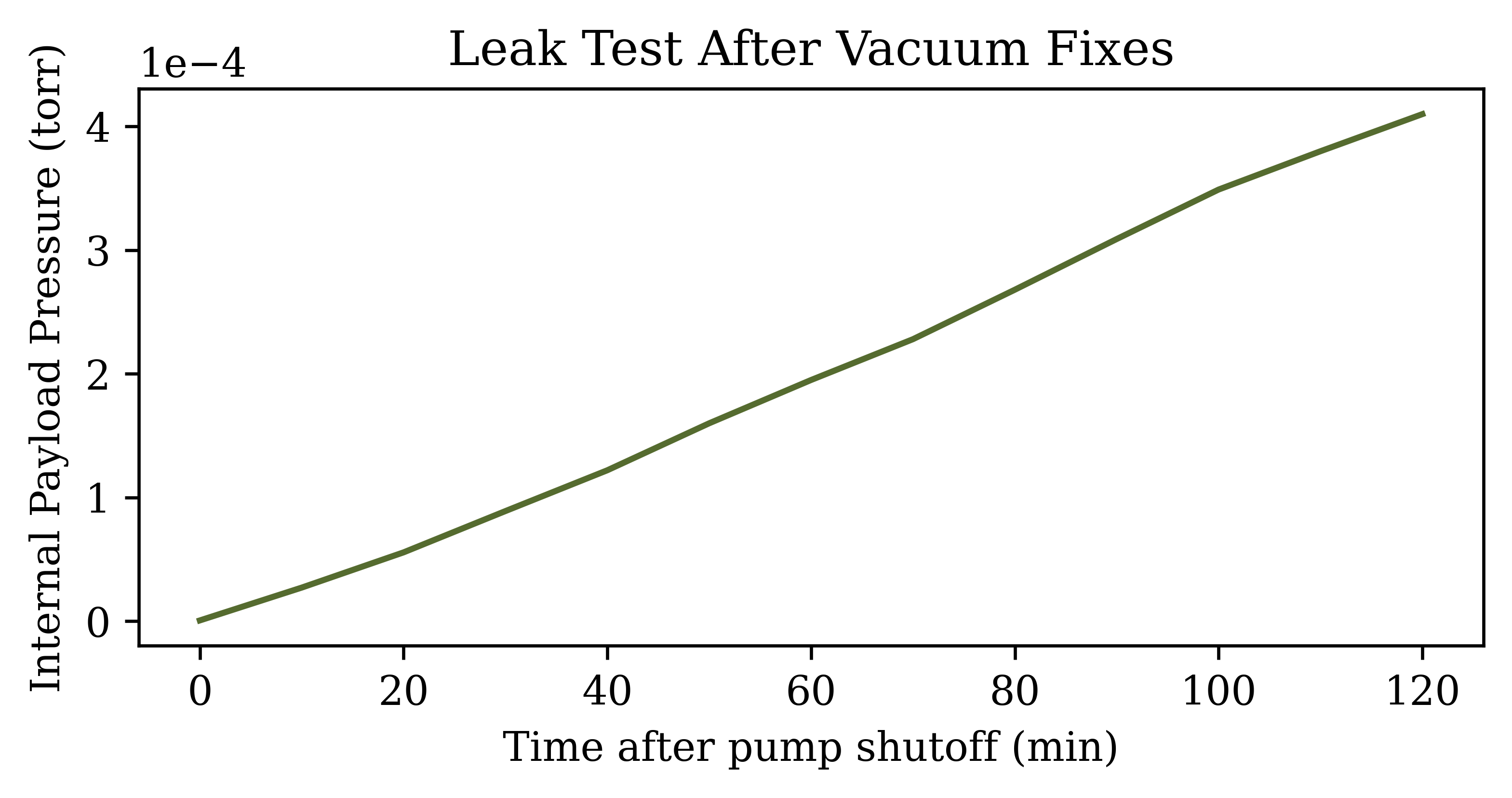}
\caption{A leak test meant to simulate launch night shows the payload on track to maintain a pressure less than 1 x 10$^{-3}$ torr over a four hour time period}
\label{fig:leak}
\end{center}
\end{figure}

A number of upgrades have been made to improve the INFUSE vacuum sealing and reduce the potential for gas saturation of the detector on future launches. Investigation of the vacuum seals post-launch determined that there were several tooling marks at the union between the adapter for the butterfly valve and the payload skins. Subsequent polishing of this area to remove the markings dropped the ultimate pressure from around 9 x $10^{-6}$ torr to 3 x $10^{-7}$ torr. In addition, different procedures are in place at the launch range to prevent the payload from sitting for too long before launch. The payload team has also made accommodations for an SAES CapaciTorr HV 200 getter pump to be installed on the detector bulkhead of the payload. A getter is a passive pump that absorbs water molecules and other contaminants; while these pumps are not designed for a volume as large as INFUSE, the capacity of the getter coupled with the improved leak rate of the payload is expected to reduce the total gas pressure in the spectrograph section to under 10$^{-3}$ torr at launch (Fig.~\ref{fig:leak}).

 \begin{figure}
\begin{center}
\includegraphics[width=4.5in]{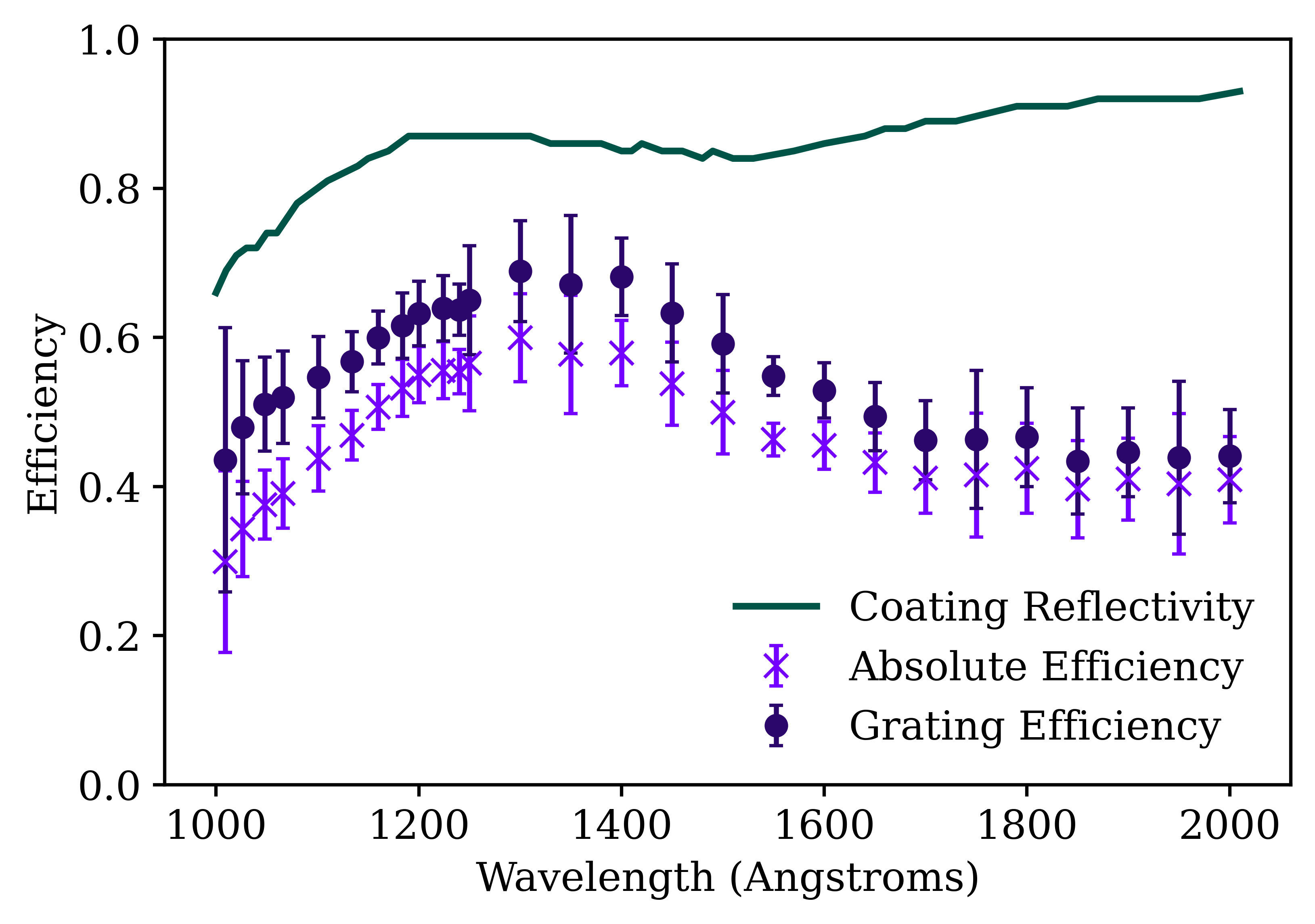}
\caption{The XeLiF coated grating has been measured in a vacuum chambers at CU Boulder by taking an incident measurement and reflected measurement from a UV light source. The absolute efficiency is the ratio of those two measurements after adjusting for dark counts. The grating efficiency was not diminished by the coating, and the absolute efficiency remains above 0.5 down to 1050 \AA. Coating reflectivity measurements are provided by GSFC and published here for the first time.}
\label{fig:XeLiF}
\end{center}
\end{figure}

\begin{figure}
\begin{center}
\includegraphics[width=\linewidth]{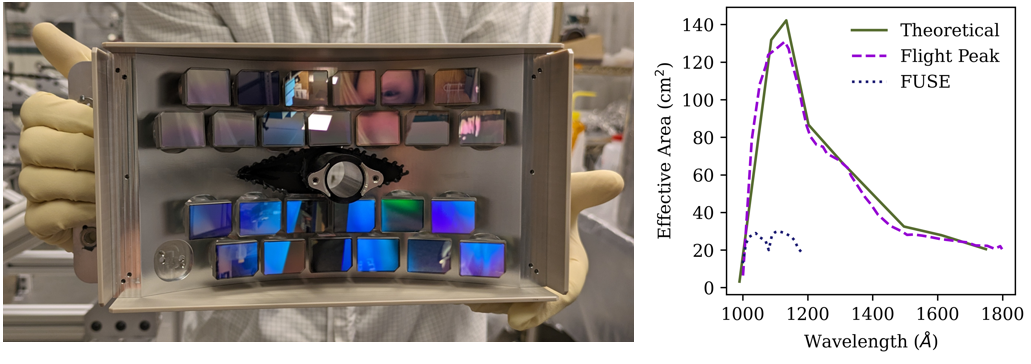}
\caption{Left: the grating array with 25 gratings installed. The 26th grating slot is for the XeLiF grating which was not ready for the first flight. Right: Theoretical and flight peak effective area of INFUSE along with the peak effective area of FUSE at launch\cite{kaiser09}. Effective area varies from grating to grating and is best near the center of the FOV; ``flight peak" is the best performing grating during the first flight based off of measuring geocoronal Ly-$\alpha$.}
\label{fig:effa}
\end{center}
\end{figure}

\subsection{Advanced UV mirror coatings for HWO}
INFUSE flies a number of advanced ultraviolet mirror coatings developed at Goddard Space Flight Center for the support of HWO; these coating advancements have helped increase the throughput and effective area of UV instruments (Fig.~\ref{fig:effa}). The primary mirror, gratings, and image slicer are coated with conventional lithium flouride (LiF) capped Al, as was used on FUSE\cite{wilkinson98}, and the secondary mirror is coated with enhanced lithium flouride capped Al (eLiF)\cite{fleming17}.  Since the initial coating of the INFUSE mirrors, a new coating option, XeLiF, has shown even more promise. Unlike eLiF, XeLiF is deposited at room temperature, making it better suited for replica gratings and large optics. In addition, early results suggest that this coating is more stable than conventional LiF\cite{quijada22,lewis24} and has improved surface roughness and uniformity. The 26th and final grating on INFUSE was set aside for the first flight so that it could be coated with XeLiF; it has now been coated with XeLiF and measured in the lab (Fig.~\ref{fig:XeLiF}) and will be flight tested during the second launch. This is the first application of this coating on a flight optic and the first deposition on a replica grating. Both conventional LiF and XeLiF perform better than magnesium fluoride + Al at wavelengths less than 120 nm, and XeLiF is now considered a leading coating candidate for HWO.

\begin{figure}
\begin{center}
\includegraphics[width=\linewidth]{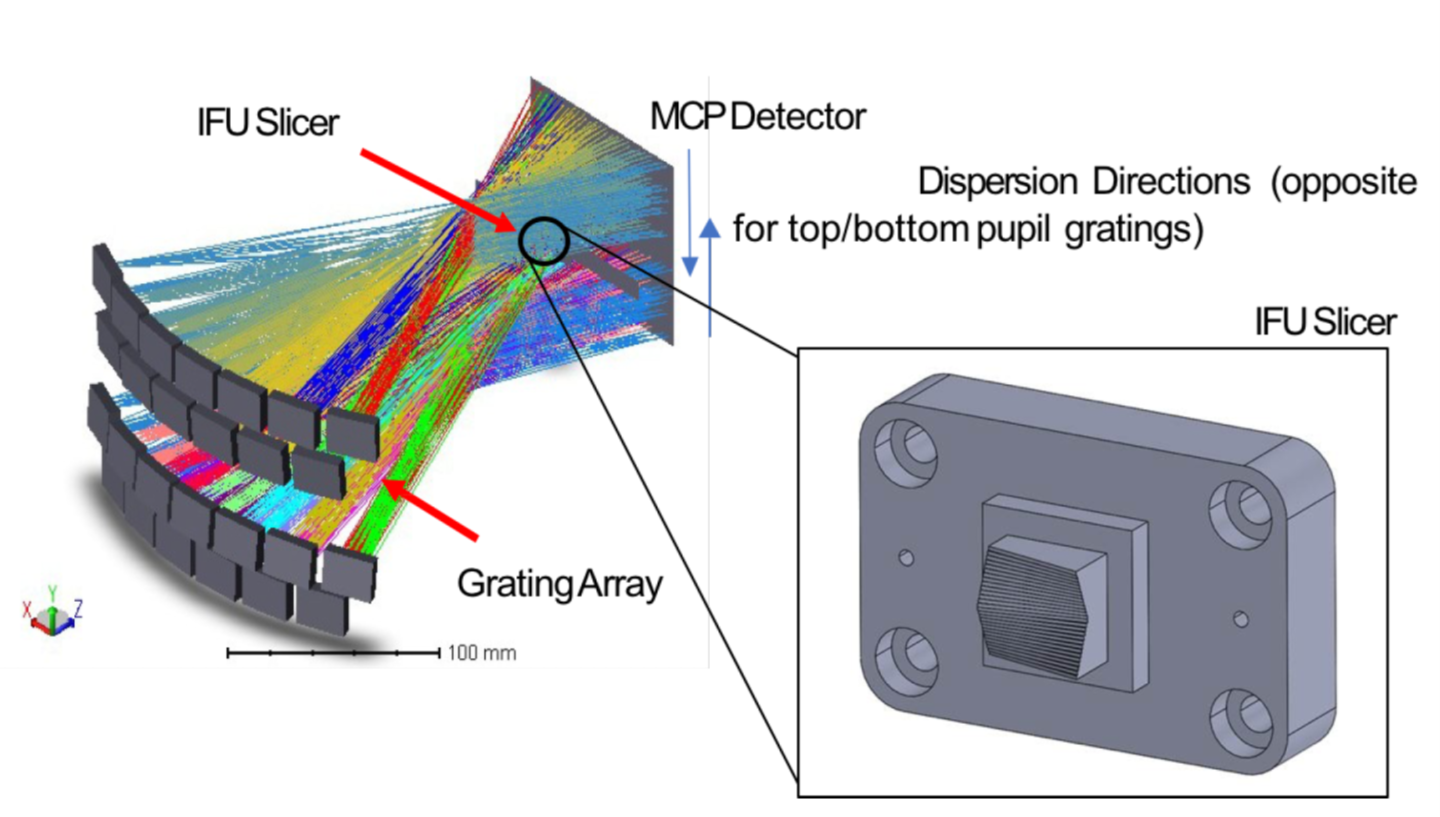}
\caption{The gratings both diffract the light and act as pupil mirrors, reducing the number of bounces and resulting in a compact design.}
\label{fig:grateray}
\end{center}
\end{figure}

\subsection{Spectrograph design for HWO and grating array alignment}
\label{sec:grate}
The design of the spectrograph section is compact, limits the number of bounces to four, and can be scaled to meet the size requirements of HWO. Light is focused onto an image slicer (Fig.~\ref{fig:grateray}), which consists of 26 slices machined out of copper-plated invar and coated with LiF + Al. These cut the image 26 ways into 5.7" by 2.57' slices, defining one dimension of spatial resolution, and direct the light to 26 identical replica holographic gratings that act as pupil mirrors. These gratings are all replica holographically ruled gratings blazed to 8.2 degrees to optimize for 1200-1400 \AA. The number of slices and gratings can be increased or decreased as needed, and while the INFUSE design minimizes the number of bounces needed, additional optics for a folded or higher resolution design would only have the effect of an $\sim$20\% loss of throughput per reflection due to the new high efficiency coatings. Science requirements for any given instrument drive this cost-benefit analysis.

\begin{figure}
\begin{center}
\includegraphics[width=3.5in]{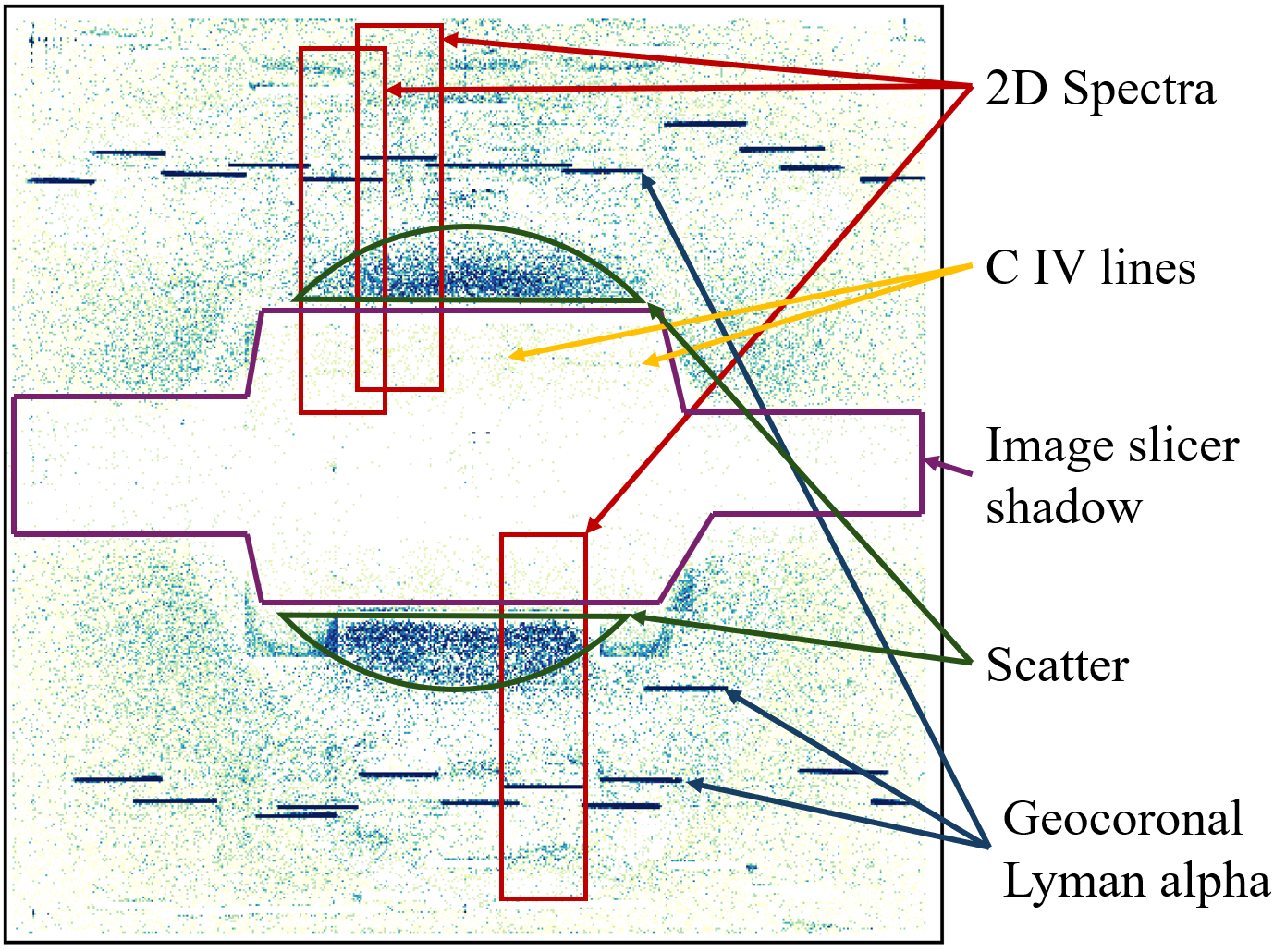}
\caption{Partially reduced image of the detector from the first INFUSE launch, which was successful at mapping the XA Region of the Cygnus Loop despite only collecting 19.4 seconds of data. The bright horizontal lines are geocoronal Ly-$\alpha$, each mapping to an element of the image slicer and representing an individual spectrogram. The central obscuration is from the image slicer and mount, and blocks scattered light but not light from the gratings - all wavelengths in the bandpass from all gratings reach the detector.}
\label{fig:data}
\end{center}
\end{figure}

For INFUSE to produce over 1,000 resolved spectra in one exposure, it must be able to fit 26 long slit spectra onto the large format detector. To this end, the spectra are designed to fall in two zigzag rows of 13 spectra each (Fig.~\ref{fig:data}). The image slicer and mount form an obscuration that protects against scattered light while allowing diffracted light from the gratings to reach the detector; no spectral coverage is lost through obscuration. Spectra are aligned such that the longer wavelengths fall towards the center of the detector, and shorter wavelengths fall closer to the edges. 

Alignment of the gratings is done by modifying sacrificial shims that attach each grating to the grating array. The grating array generally sets the angle between the gratings, image slicer, and detector, and the shims are designed to be modified in order to fine tune the angle and focus location. Due to a compressed schedule to meet the launch window, only one iteration of three of the sacrificial shims was taken to improve alignment for the first flight, resulting in several partially overlapping spectrograms. This was acceptable for the first flight, as the Cygnus Loop is an emission-line source and the risk of spectral confusion was minimal; NGC 2366 is a continuum source, however, and requires improved alignment. Realignment of the grating will occur in an iterative manner, with the spectrogram locations noted and new sacrificial shims machined to adjust towards the nominal position.

\subsection{Aspect camera and fold mirror improvements}
\label{sec:aspect}
An aspect camera pointed at a fold mirror around the INFUSE image slicer guides navigation during flight; however, the FOV of the aspect camera and the area occupied by the image slicer combine to limit the usable area on the fold mirror. The fold mirror was also hand polished for the first flight to limit the magnitude of visible stars. During the first flight, the guide star could not be seen by the aspect camera in the allotted 115 second interval, so the timed backup from the attitude control system (ACS) took INFUSE to the science target. Even with the aspect camera issues, we achieved pointing of about 20" and jitter of less than 2". For the second flight, the fold mirror will hug the image slicer more closely, and the FOV of the aspect camera system will be increased from 10' x 4' to 12' x 6'. It will also be professionally polished. The combination of these three changes will improve guidance during flight, and should increase time on target and make calibrating the location of flight data easier.

\subsection{Control of scattered light}
\label{sec:scatter}
The first flight of INFUSE revealed moderate scattered light from geocoronal Lyman alpha (Ly-$\alpha$) grazing off of smooth surfaces in the telescope baffles. This was expected, as the INFUSE detector lies directly behind the telescope focus, so small angle stray light not intercepted by the baffles of the spectrograph yields a diffuse glow around the image slicer (Fig.~\ref{fig:data}). A future design for HWO or an orbital surveyor could benefit from a folded optical system to remove the direct line-of-sight from the telescope to the detector; this was not deemed favorable for the INFUSE design, as throughput and instrument compactness were more favorable for a sounding rocket design.

Scattered light during the first flight matched pre-flight predictions. An estimated 2,000 Rayleighs of geocoronal Ly-$\alpha$ at midnight translates to 1.59 x 10$^{8}$ photons s$^{-1}$ cm$^{-2}$  sr$^{-1}$. The spectrograph aperture has an FOV of about 20', with about 215 cm$^{2}$ of effective area at that point from the primary and secondary; this results in about 3.6 million photons per second of Lyman alpha with the potential to scatter onto the detector. For on-axis light, the FOV is much smaller, and about 8,000 cts per second on the detector of geocoronal Ly-$\alpha$ were expected after bouncing off the image slicer and grating. The recorded count rate was closer to 20,000 counts per second, with 12,000 counts per second of scattered light concentrated around the central obscuration. For comparison, the dark rate as measured in the lab across the whole detector is $\sim$ 16 counts per second.

While the amount of light met expectations, given the short exposure time (\S~\ref{sec:vac}), the scattered light reduced the signal-to-noise ratio of some of the observed spectral lines. The geometry of the spectrograph is such that second-order geocoronal Ly-$\alpha$ from one spectrogram falls in the corresponding spectrogram across the image slicer obscuration (Fig.~\ref{fig:data}) right near where first order O VI lands. In addition to the scatter, this second order Ly-$\alpha$ also contaminated flight data, making extraction of O VI in particular more difficult.

To solve these problems for the second flight, the spectrograph baffle will be modified, which should suppress the second-order Ly-$\alpha$, and the baffle on the image slicer bridge will be extended to block a larger portion of the central region as well. In addition, parts of the spectrograph section will be iridated to reduce the amount of scattered light.






\section{The Spectroscopic Ultraviolet Multi-Object Observatory}
\label{sec:SUMO}

\begin{figure}
\begin{center}
\includegraphics[width=\linewidth]{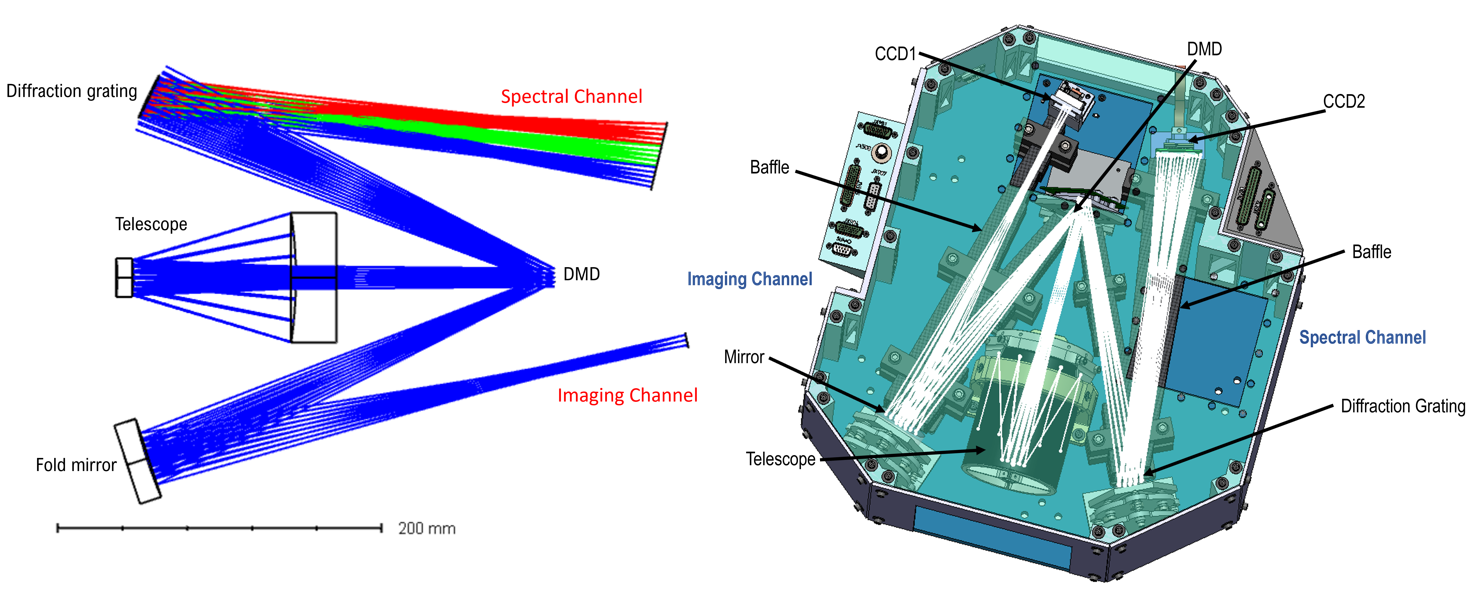}
\caption{Ray trace and 3D CAD of the SUMO secondary payload, which will observe with DMDs in space for the first time as a ride along on the second launch of INFUSE.}
\label{fig:SUMO}
\end{center}
\end{figure}

\begin{figure}
\begin{center}
\includegraphics[width=4in]{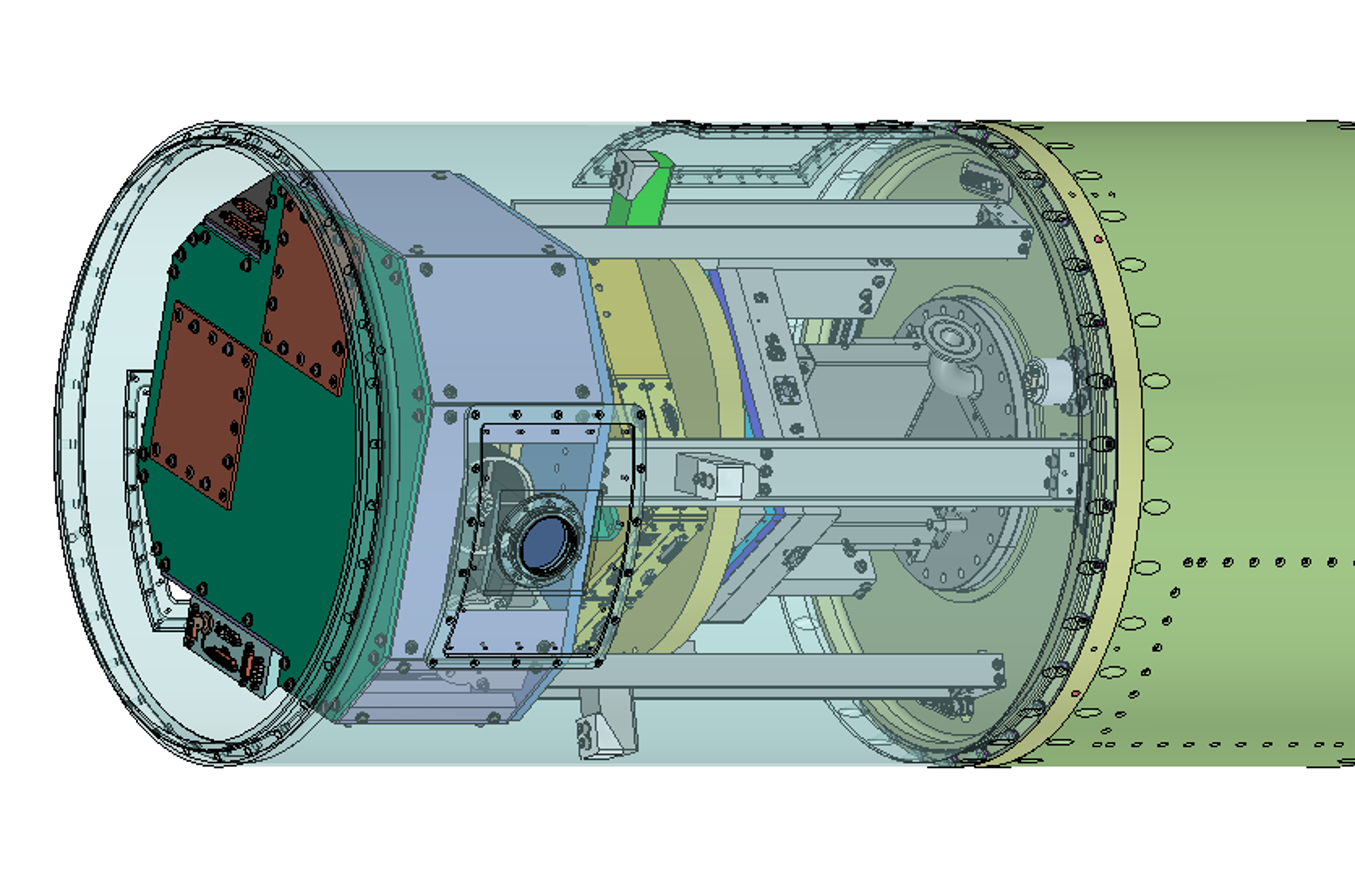}
\caption{SUMO sits at the forward end of the INFUSE payload on top of the electronics section.}
\label{fig:sumoinf}
\end{center}
\end{figure}

INFUSE will also carry SUMO, a ride along instrument demonstration. SUMO is a mission concept designed for small satellite platforms, built around an advanced DMD-based spectrograph. DMDs are arrays of tiny, tilt-controlled mirrors that have potential as an alternative path to multi-object spectroscopy. They are inexpensive, reliable, and reflective, and typically have 1024 x 768 or 2048 x 1024 mirrors with bi-stable orientations (+12° or -12° tilt). This gives the bi-channel shape with the imaging and spectral sections of the instrument (Fig.~\ref{fig:SUMO}). As part of the SUMO technology maturation program, the SUMO prototype will be developed and deployed on INFUSE in Fall 2025 (Fig.~\ref{fig:sumoinf}).

Space-based multi-object spectroscopy is an enabling technology for a wide range of astrophysics research programs, including efforts to understand the processes of star formation and galaxy evolution. Because the spectroscopic diagnostics of these processes exist in the UV range and are often distributed across an extended angular area, there is a need for high-efficiency object selection technology for UV multi-object spectrometers. The prototype design of SUMO is part of a larger concept and can be scaled to eventually suit a range of missions. SUMO is designed for astrophysics research in the near and far UV ranges. This will be the first time a DMD-based instrument is deployed in space; however, DMDs have been tested repeatedly to show that they are not particularly susceptible to the shock and vibration environments of launch or the radiation effects of orbit.

The telescope is a custom Cassegrain telescope developed by Nu-Tek Precision Optical Corporation. It has an 8 cm diameter parabolic primary mirror and a 24 mm diameter hyperbolic secondary mirror. The instrument will also consist of a custom radiation-tolerant controller. SUMO is a secondary payload, which means it needs to be integrated successfully into the INFUSE sounding rocket without interfering with its primary mission. For this reason, SUMO has autonomous operation. The SUMO prototype is attached to C-channel brackets along the walls of the rocket skin above the existing electronics deck and observes through a window in the side of the rocket skin. Current estimates for the instrument are a mass of 40 lbs and dimensions of 415.63 x 456.35 x 190.96 mm. The current design consists of the Cassegrain telescope, a custom space-suitable DMD controller, two detectors, diffraction grating, fold mirror, flight computer, EPS board, along with the DMD and carrier board. The near-UV detector has heritage from the Colorado Ultraviolet Transit Experiment\cite{nell21}, the first astrophysics CubeSat, and was also developed at CU Boulder. For more information see ``Development, building, and testing of the Spectroscopic Ultraviolet Multi-Object Observatory (SUMO) prototype for deployment on the INFUSE sounding rocket" by Halferty et al.\cite{halferty24}

\section{Potential Adaptation for Habitable Worlds Observatory}
\label{sec:HWO}
The INFUSE design is scalable to larger instruments, and is meant as a pathfinder for a  potential secondary channel on HWO, similar to the JWST-MIRI IFS. As with INFUSE, the spatial and spectral resolutions and FOV are primarily functions of the amount of detector real estate available for packing spectrograms. We hypothesize a potential IFS for HWO by assuming the following:

\begin{itemize}
    \item There are two 100 x 100 mm MCP detectors located on either side of the image slicer, controlling stray light better than the single detector approach of INFUSE. Each detector has 20 $\mu$m resolution. 
    \item The image slicer is located at the telescope focal plane. In reality, this may not be possible depending on the HWO instrument design, in which case the beam will have to be picked off and re-focused within the instrument volume, reducing throughput. 
    \item As a supporting instrument, all optics must be fixed in position. While putting either the image slicer or gratings on a mechanism for selecting different channels would be ideal, this would require additional budget. We therefore assume that the optimal design is one that spans the 1000 - 1600 \AA\ bandpass, covering key spectral lines like O VI, \lya , and C IV; this bandpass may change depending on the driving science.
    \item Following the Exploratory Analytic Case telescope architectures\cite{eacs}, we assume an F/18, 6 m diameter telescope beam, for a plate scale of $\sim$ 1.9\asec\ mm$^{-1}$. 

    \end{itemize}

Given these assumptions, we can estimate some of the instrument specifications. While the final balance of FOV and resolution is ultimately a function of the science requirements, the spectral resolution for the bandpass we selected is set by the detector width and resolution itself. For a 600 \AA\ spectrum diffracted and focused over 100 mm of detector length, the dispersion is 6 \AA\ mm$^{-1}$. At a 20 $\mu$m resolution, this corresponds to a best-case point-source resolution of 0.12 \AA , or a resolving power of 10,000 at \lya . In reality, it may be challenging to control spectral aberrations to this degree over a large IFS, therefore R $\sim$ 5000 is likely a more realistic average. 

While there exists no specific science guidance for angular resolution for an IFS, the early MOS studies suggest an appropriate objective of angular resolution $<$ 100 mas. Given our estimated plate scale of 1.9\asec\ mm$^{-1}$, achieving this resolution on a diffuse source sets the image slicer slit width to $\approx$ 50 $\mu$m, with the maximum cross-dispersion resolution 20 $\mu$m $\times$ 1.9\asec\ mm$^{-1}$ $\approx$ 37 mas. If the 50 $\mu$m slice is filled with a diffuse source, the maximum spectral resolving power will decrease to R $\sim$ 4000 (10,000 x (20 $\mu$m/50 $\mu$m)). 

\begin{figure}
\begin{center}
\includegraphics[width=3in,trim= 10 10 10 10,clip]{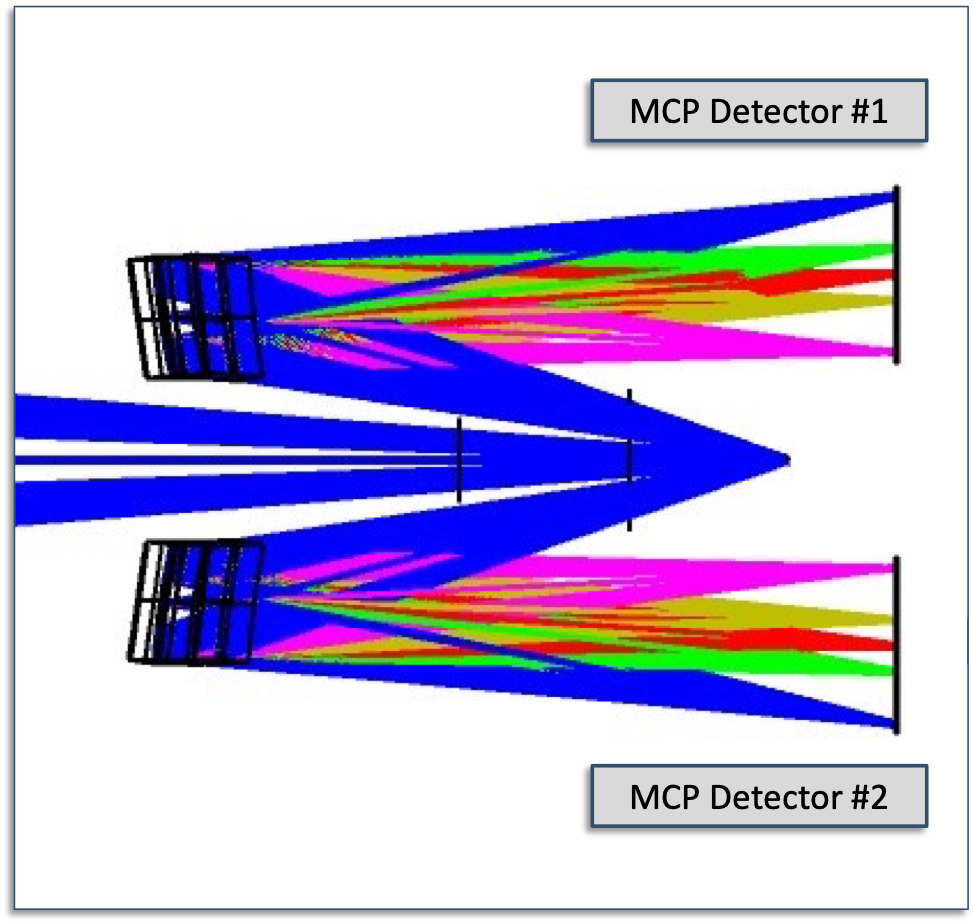}
\caption{A ray trace of a potential IFS design to fit in HWO architecture.}
\vspace{-20pt}
\label{fig:hwoifu}
\end{center}
\end{figure}

The FOV then becomes a function of how many spectrograms can fit on each detector and how many gratings can be arrayed. We set objectives to have a square FOV and to maximize the cross-dispersion resolution. For a total detector width of 200 mm (2$\times$ 100 mm detectors), this corresponds to roughly 30 slices per detector (assuming a small gap between each spectrogram), or 60 total, each 3.15 mm long, subtending a 6\asec\ $\times$ 6\asec\ FOV. 

Such a design is similar to the INFUSE design, which features an F/16 telescope and 26 gratings imaging onto a 94 $\times$ 94 mm detector. The final instrument would cover major FUV spectral features at moderate/low resolving powers of 4000 for diffuse gas and up to 10,000 for point sources over a FOV commensurate with low redshift (z $\sim$ 0.1) galaxies at 100 mas resolution. $sim$ 7,000 individual 1D spectra would be recorded per observation, providing excellent efficiency for mapping extended sources. 

With additional investment, the slicer could be placed on a mechanism and a variety of different slicer could be made, trading increased FOV for decreased spatial or spectral resolving power. For example, a slicer with thicker, 200 mas-wide slices would increase the FOV to 6\asec\ $\times$ 12\asec\ at half the spectral resolving power. It would be more challenging to swap the grating arrays, but hybrid designs, such as with one grating array covering short wavelengths and the other longer wavelengths at higher spectral resolution, are possible depending on the definition of the science objectives.The ideal design would also give access to the He II 1640, O III] 1661, 1663 and C III] 1907, 1909 spectral lines. An INFUSE-like IFS could be added as a supporting instrument to a more multi-purpose MOS on HWO.

\section{Conclusion}
\label{sec:conc}

INFUSE demonstrated the validity of large format XS MCP detectors, eLiF grating coatings, and UV-optimized integral field spectroscopy in the FUV during its flight on October 29th, 2023. These technologies are ready to advance to larger missions, and an IFU could be added to HWO in a space-efficient manner. IFUs are incredibly time-efficient as well; on the first flight of INFUSE, data surpassing previous attempts by FUSE and IUE was collected in just 19.4 seconds, providing spectral line mapping across the Cygnus Loop XA region. INFUSE will fly again in Fall 2025 with a better guide camera FOV, less scattered light, improved protection from range anomalies, a XeLiF coated grating, a realigned grating array, and the ride along SUMO demonstrating a new form of MOS.

\acknowledgments  
This work would not be possible without the contributions of many individuals not listed as authors on the paper. In particular, special thanks to Jack Williams and Diane Brening for electrical work and Dana Chafetz for the mechanical design, and to Nicholas Kruczek for helping take the XeLiF measurements. The research presented in this article was funded by NASA Grant NNH21ZDA001N-APRA. A version of this article originally appeared as an SPIE Conference Proceedings\cite{haughton24}.
\newline

\noindent\textit{Code and Data Availability Statement}
\newline
\noindent The data presented in this article are publicly available in GitHub at DOI: 10.5281/zenodo.14736591\\

\noindent\textit{Disclosure Statement}
\newline
\noindent The authors declare there are no financial interests, commercial affiliations, or other potential conflicts of interest that have influenced the objectivity of this research or the writing of this paper.

\bibliography{report} 
\bibliographystyle{spiejour} 

\end{document}